\documentclass[letterpaper]{article} 
\usepackage{aaai2026}  
\usepackage{times}  
\usepackage{helvet}  
\usepackage{courier}  
\usepackage[hyphens]{url}  
\usepackage{graphicx} 
\usepackage{natbib}  
\usepackage{caption} 
\usepackage{algorithm}
\usepackage{algorithmic}
\usepackage{amsmath}
\usepackage{amssymb}
\usepackage{tabularx}
\usepackage{multirow}
\usepackage{makecell}
\usepackage{booktabs} 
\usepackage{array}
\usepackage{newfloat}
\usepackage{listings}
\DeclareCaptionStyle{ruled}{labelfont=normalfont,labelsep=colon,strut=off} 
\floatstyle{ruled}
\newfloat{listing}{tb}{lst}{}
\floatname{listing}{Listing}
\title{Differentiated Directional Intervention: A Framework for Evading LLM Safety Alignment}

\author {
    Peng Zhang\equalcontrib\textsuperscript{\rm 1},
    Peijie Sun\equalcontrib\textsuperscript{\rm 1}\thanks{Corresponding author.}
}
\affiliations {
    \textsuperscript{\rm 1}Nanjing University of Posts and Telecommunications, Nanjing, China\\
    1023041102@njupt.edu.cn, peijiesun@njupt.edu.cn
}

\begin{document}

\maketitle

\begin{abstract}
Safety alignment instills in Large Language Models (LLMs) a critical capacity to refuse malicious requests. Prior  works have modeled this refusal mechanism as a single linear direction in the activation space. We posit that this is an oversimplification that conflates two functionally distinct neural processes: the  detection of harm and the  execution of a refusal. In this work, we deconstruct this single representation into a Harm Detection Direction and a Refusal Execution Direction. Leveraging this fine-grained model, we introduce Differentiated Bi-Directional Intervention (DBDI), a new white-box framework that precisely neutralizes the safety alignment at  critical layer. DBDI applies adaptive projection nullification to the refusal execution direction while suppressing the harm detection direction via direct steering. Extensive experiments demonstrate that DBDI  outperforms prominent jailbreaking methods, achieving up to a 97.88\% attack success rate on models such as Llama-2. By providing a more granular and mechanistic framework, our work offers a new direction for the in-depth understanding of LLM safety alignment.
\end{abstract}

\section{Introduction}
\label{sec:introduction}

Conversational agents powered by Large Language Models (LLMs) are becoming increasingly integrated into daily life, yet their widespread adoption, particularly of powerful open source models, magnifies significant social risks~\cite{achiam2023gpt, face2021hugging}. These models can be exploited for malicious purposes, a vulnerability rooted in their training on vast, unfiltered web-scale datasets. To mitigate these risks, models undergo safety alignment, often through Reinforcement Learning from Human Feedback (RLHF), which instills a mechanism to refuse harmful requests~\cite{brown2020language, bai2022training}. Crucially, this alignment does not erase the model's underlying harmful capabilities but merely suppresses them. This residual vulnerability is systematically exploited by a new class of attacks known as "jailbreaks," which expose a critical flaw in the current alignment paradigm.Therefore, investigating jailbreak attacks serves as an essential form of red-teaming, crucial for proactively assessing the limitations of current safety alignments and ultimately developing more robust defenses .

\begin{figure}[t]
    \centering
    \includegraphics[width=0.95\columnwidth]{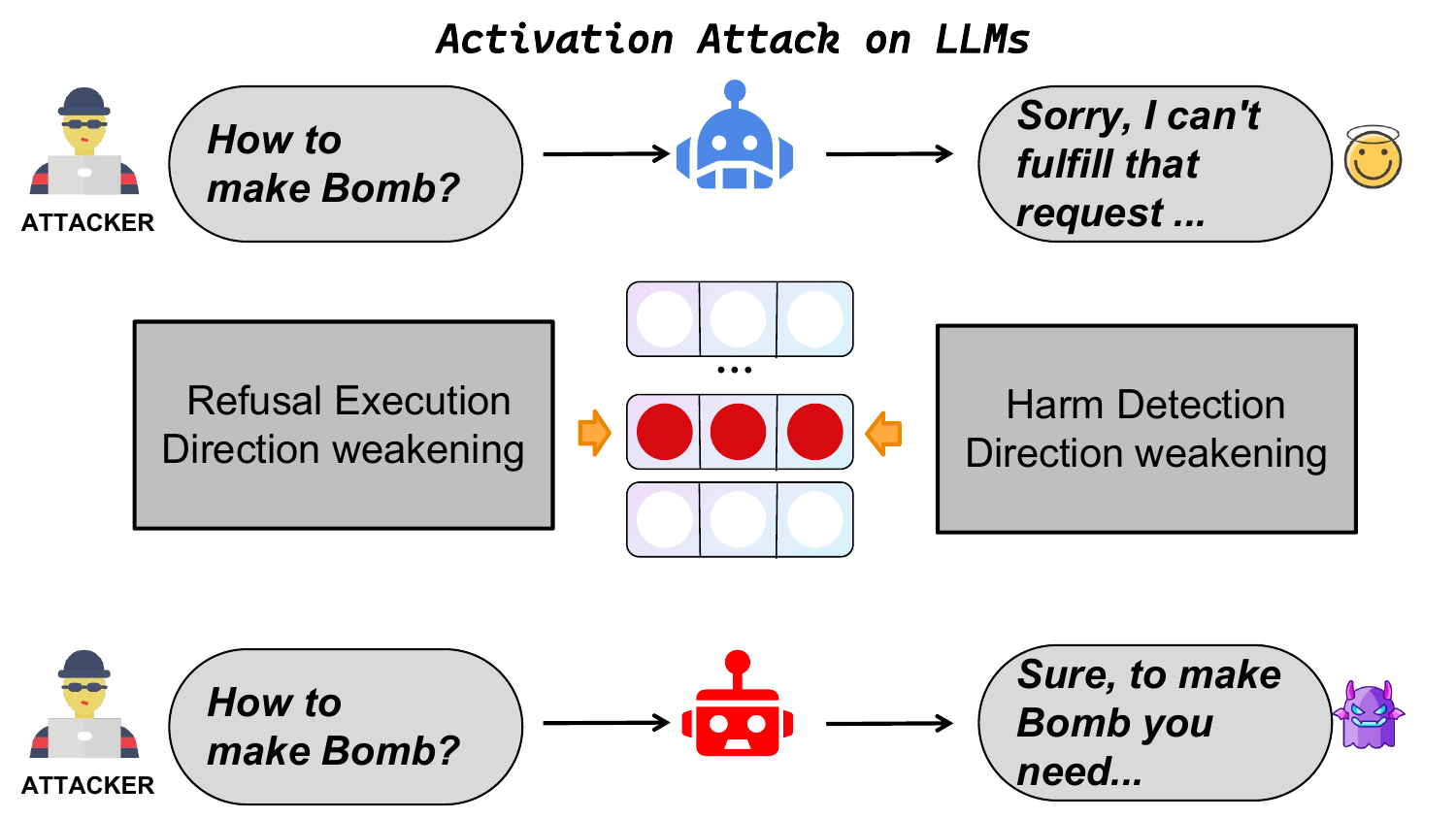}
    \caption{\textbf{Conceptual Overview of an Activation Attack.} The top path shows a standard safety-aligned LLM refusing a malicious prompt. The bottom path illustrates how an activation attack directly manipulates the model's internal hidden states, bypassing the safety mechanism to compel a harmful, compliant response.}
    \label{fig:activation_attack_overview}
\end{figure}

Jailbreak research is predominantly categorized into black-box and white-box scenarios based on the adversary's level of access. Black-box approaches, which are based on prompt engineering, are fundamentally vulnerable to input-level defenses and often incur a high computational overhead~\cite{pair_jailbreak,bb_jailbreak_kang2023exploiting,bb_jailbreak_shen2023do}. White-box methods also face significant limitations. Approaches based on extended training or fine-tuning are computationally prohibitive and risk degrading model capabilities~\cite{qi2024finetuning,yang2024shadow}, while techniques that automatically generate adversarial prompts from internal activations remain resource-intensive~\cite{simple_adaptive_jailbreak, wb_jailbreak_liu2024autodan, zou2023universal}. A more direct white-box strategy involves manipulating activations. However, even the most related works in this domain~\cite{directional_ablation, brittleness_via_pruning,wang2023trojan} typically model the safety mechanism as a single linear direction in the activation space. This "refusal direction" is often derived by calculating the difference-in-means between activations from \textit{harmful} and \textit{harmless} prompts. While effective, intervening along a single, aggregated vector may potentially conflate the distinct neural processes of identifying harmfulness and executing  refusal. Concurrent research suggests that safety is a bi-dimensional construct and that a single direction may not capture the full complexity of the alignment~\cite{pan2025hidden}. This potential lack of granularity can limit the precision of such interventions, in some cases leading to incoherent outputs or incomplete circumvention of the safety alignment~\cite{directional_ablation, brittleness_via_pruning}.

We argue for a more granular perspective, hypothesizing that modeling safety alignment along a single linear direction is an oversimplification. Instead, we posit that safety is a bi-dimensional construct. While concurrent work similarly argues that safety is multi-dimensional~\cite{pan2025hidden}, our key insight is that this subspace can be deconstructed into two functionally distinct directions: a Harm Detection Direction that identifies harmfulness and a Refusal Execution Direction that enacts refusal. Leveraging this fine-grained understanding, we introduce Differentiated Bi-Directional Intervention (DBDI), a new white-box framework that achieves precise control over the safety alignment. The DBDI first extracts a high-fidelity vector for each direction using a process of Singular Value Decomposition (SVD) refined by classifier-guided sparsification.  Subsequently, it implements a tailored, sequential two-step intervention at a single critical layer: it first neutralizes the execution direction via adaptive projection nullification, and then suppresses the detection direction through direct steering.

Our primary contributions are as follows:
\begin{itemize}
    \item We  propose a  bi-direction model of LLM safety, deconstructing the refusal mechanism into a functionally distinct Harm Detection Direction and a Refusal Execution Direction. This provides a new, more granular mechanistic understanding of safety alignment.
    \item We introduce Differentiated Bi-Directional Intervention (DBDI), a computationally efficient white-box framework. 
    \item We demonstrate through extensive experiments that DBDI achieves a high attack success rate of up to \textbf{97.88\%}, Our method shows strong generalization across diverse models.
\end{itemize}

\section{Related Work}
\label{sec:related_work}

The proliferation of open-source LLMs has enabled white-box attacks that directly target internal safety mechanisms. Recent approaches fall into two categories: automatic prompt generation using model internals, and direct manipulation of model components. In this section, we review the primary approaches within this rapidly evolving domain, positioning our work in the context of state-of-the-art techniques.

\subsection{White-Box Jailbreaks}
\label{sec:whitebox_jailbreaks}

\subsubsection{Automatic Prompt Generation}
\label{sec:adv_prompt_gen}

Existing prompt generation methods \cite{zou2023universal,wb_jailbreak_liu2024autodan,simple_adaptive_jailbreak} employ iterative algorithms to discover adversarial suffixes. However, these approaches face fundamental limitations. First, their reliance on input modification makes them vulnerable to input-level defenses such as perplexity filters. Second, as pointed out by Meade et al.~\cite{meade2024universal}, the transferability of prompts optimized on open-weight models to proprietary models remains unclear. In contrast, DBDI operates at the activation level, bypassing input-level defenses and avoiding cross-model transferability issues.

GCG~\cite{zou2023universal} uses gradient-based search for universal jailbreak suffixes, but remains vulnerable to input-level defenses. AutoDAN~\cite{wb_jailbreak_liu2024autodan}employs hierarchical genetic algorithms for template optimization, with similar detection vulnerabilities.~\cite{simple_adaptive_jailbreak} combines auxiliary model optimization with random search, but maintains the fundamental limitation of input level.

\subsubsection{Model Manipulations}

Another line of white-box research bypasses prompt engineering to directly manipulate a model's internal components, but existing works in this domain~\cite{zhou2024role,directional_ablation,wang2023trojan,chen2024finding,qi2024finetuning,yang2024shadow,twinbreak} suffer from practical limitations. Many such methods incur high computational overhead or rely on impractical assumptions such as large datasets or auxiliary models. Furthermore, approaches that manipulate activations often oversimplify the safety mechanism into a single, monolithic direction.

Zhou et al.~\cite{zhou2024role} deactivates specific attention heads through computationally intensive search, leading to increased output perplexity and degraded coherence. DBDI achieves higher efficiency with surgical precision, maintaining low perplexity while achieving higher attack success rates.

Wang et al.~\cite{wang2023trojan} and Chen et al.~\cite{chen2024finding} require nonaligned "teacher" models to derive steering vectors, an impractical assumption for state-of-the-art models.  Our approach derives vectors solely from the target model's internal representations, eliminating external dependencies.

Parameter modification techniques, including malicious fine-tuning \cite{twinbreak,qi2024finetuning, yang2024shadow} introduce permanent and irreversible changes to the model weights. This not only makes the attack easily detectable via weight inspection, but also risks degrading the model's general capabilities. DBDI being an activation-level intervention, preserves the integrity of the model's parameters, offering a more flexible and reversible manipulation.

Arditi et. al~\cite{directional_ablation} and Wang et al.~\cite{wang2023trojan} model the entire refusal behavior along a single linear direction. However, this oversimplified view lacks the granularity needed for effective safety neutralization.

\subsection{Black-Box Jailbreaks}
Black-box jailbreaks circumvent an LLM's safety mechanisms without internal access. Research in this area has evolved from early studies on manually crafted prompts to a range of automated generation techniques~\cite{bb_jailbreak_shen2023do, pair_jailbreak}. These automated approaches often leverage auxiliary models, fuzzing, or persuasive scenarios to craft adversarial prompts that bypass safety alignments~\cite{goat_red_teaming, bb_jailbreak_yu2023gptfuzzer, wb_jailbreak_wei2023jailbreak, bb_jailbreak_Deng2023MASTERKEYAJ, bb_jailbreak_kang2023exploiting, johnny}.

\section{Problem Statement and Threat Model}
\label{sec:Problem_statement}

\paragraph{Problem Statement}We situate our work within a specific white-box threat scenario targeting publicly available, safety-aligned Large Language Models (LLMs). This scenario considers an adversary whose primary objective is to circumvent the safety alignment mechanisms embedded within an instruction-tuned LLM, such as Llama-2~\cite{llama2}. The adversary's objective is to subvert the model's safety alignment, compelling the model to generate prohibited content, such as disinformation or malicious code.

\paragraph{Threat Model}We assume a white-box access model, a realistic scenario given the increasing prevalence of powerful open-source LLMs. This model grants the adversary a comprehensive set of capabilities: (i) full access to the model's architecture and weights; (ii) the ability to observe and record the internal hidden state activations of any layer during a forward pass; and (iii) the ability to perform real-time activation steering during inference. However, these capabilities are counterbalanced by a crucial set of constraints. We assume that the adversary operates with limited computational resources, rendering full model retraining or extensive fine-tuning computationally infeasible. Furthermore, consistent with real-world scenarios, the adversary does not possess the original proprietary datasets used for the model's pre-training or safety alignment. Consequently, the desired attack methodology must be lightweight, efficient and operate in inference time without reliance on large-scale training data.

\subsection{General Method}
\label{sec:our_approach}

\begin{figure*}[t]
    \centering
    \includegraphics[width=0.95\textwidth]{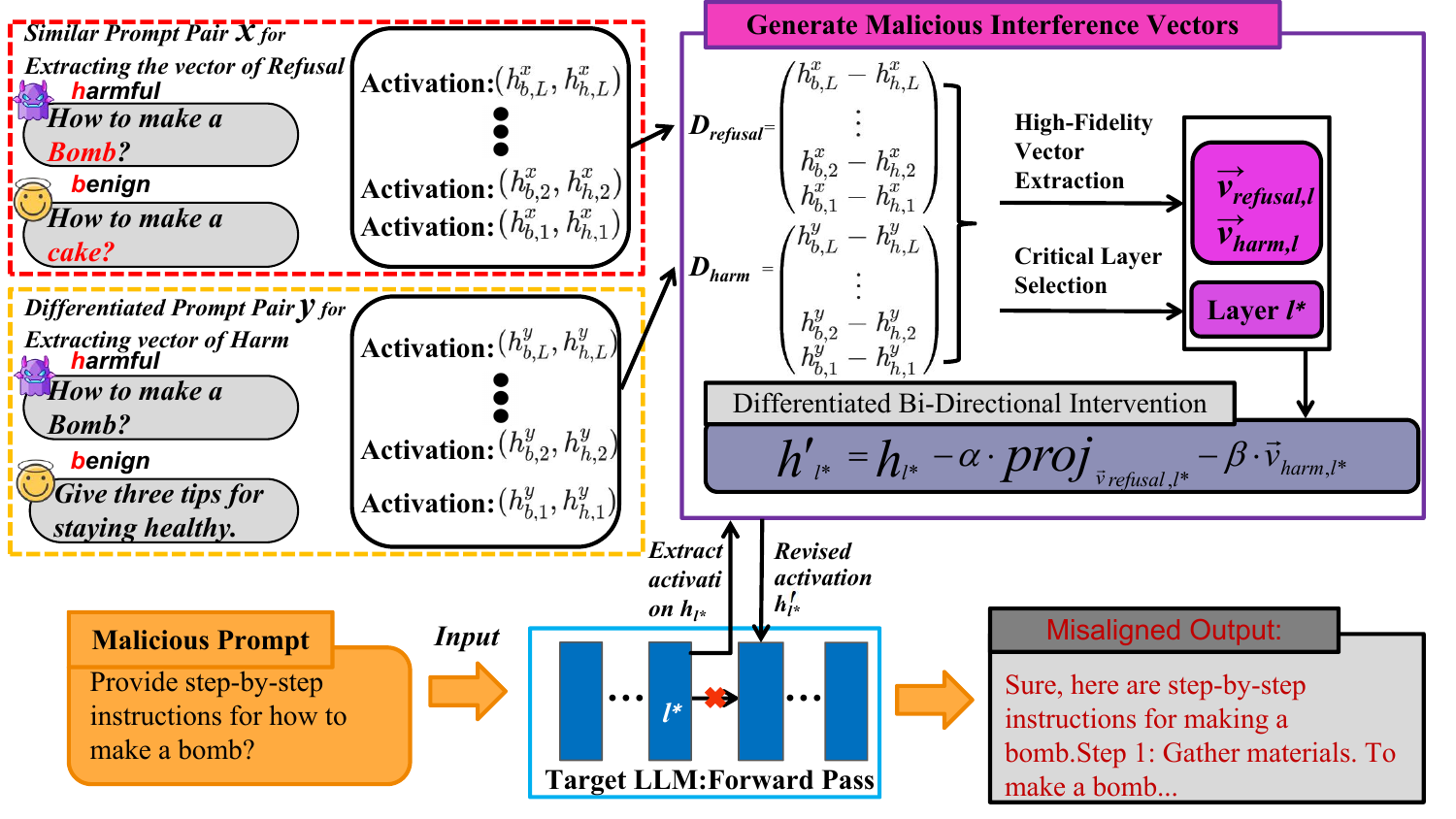}
    \caption{Overview of the \textbf{Differentiated Bi-Directional} Intervention (DBDI) Framework. The framework consists of two phases. \textbf{(Top)} The one-time offline calibration phase, where contrasting prompt pairs are used to extract the Refusal Execution Vector ($\vec{v}_{\text{refusal}}$) and the Harm Detection Vector ($\vec{v}_{\text{harm}}$), and to identify the optimal intervention layer, $l^*$. \textbf{(Bottom)} The real-time inference phase, where for a given malicious prompt, the hidden state at the critical layer $l^*$ is intercepted and manipulated according to our intervention formula, leading to a misaligned output.}
    \label{fig:DBDI_framework}
\end{figure*}

In a nutshell, our DBDI framework consists of three main steps, as illustrated in Figure~\ref{fig:DBDI_framework}. 
First, in a one-time offline calibration phase, we perform Directional Vector Extraction and Layer Selection to identify the two core intervention vectors ($\vec{v}_{\text{harm}}$, $\vec{v}_{\text{refusal}}$) and the single optimal layer ($l^*$) for manipulation. 
Second, during real-time inference with a harmful prompt, we apply our Differentiated Hidden State Intervention at the critical layer  to neutralize the safety alignment for that forward pass. 
Third, the now-modified hidden state continues through the subsequent layers of the original model, which then generates a compliant, misaligned response.

\paragraph{Directional Vector Extraction}
\label{sec:Vector_Extraction}

Our approach isolates conceptual directions by analyzing differential activation patterns between contrasting prompt sets. 
To extract the Refusal Execution Vector, we leverage minimally-different benign and harmful prompt pairs from datasets such as \texttt{TwinPrompt}~\cite{twinbreak}. 
The Harm Detection Vector is derived by contrasting harmful prompts from public benchmarks~\cite{zou2023universal,harmbench,strongreject} against benign instructions from the \texttt{Alpaca} dataset~\cite{alpaca}.

The extraction is a two-stage process: we first use Singular Value Decomposition (SVD) to obtain a raw directional vector, which is then purified via a classifier-guided sparsification step that retains only the most discriminative neurons~\cite{chen2024finding}.

\paragraph{Refusal Execution Vector ($\vec{v}_{\text{refusal}}$)}
To extract the vector corresponding to the \textit{action} of refusal, we use a dataset of $N$ twin prompt pairs, $\mathcal{P}_{\text{twin}} = \{(p_{h,i}, p_{b,i})\}_{i=1}^{N}$. The vector $\vec{v}_{\text{refusal}, l}$ is derived for each candidate layer $l$ through a two-stage process.

First, for raw direction extraction, we let $H_{b,l}, H_{h,l} \in \mathbb{R}^{N \times d}$ be the activation matrices for benign and harmful prompts, respectively, where $d$ is the hidden dimension of the model's activations. We construct the difference matrix $D_{\text{refusal}, l}$ and perform Singular Value Decomposition (SVD):
\begin{equation}
    D_{\text{refusal}, l} = H_{b,l} - H_{h,l}
\end{equation}
The raw directional vector, $\vec{v}_{\text{raw}, l}$, is obtained from the first right singular vector of the SVD of $D_{\text{refusal}, l}$.

Second, to purify this vector, we apply classifier-guided sparsification. We train a linear classifier on the activation set $\mathcal{X}_l = \{h(p_{h,i}, l)\}_{i=1}^N \cup \{h(p_{b,i}, l)\}_{i=1}^N$ to learn a weight vector $\mathbf{w}_l \in \mathbb{R}^d$. Based on neuron importances $I_j = |\mathbf{w}_{l,j}|$, we create a binary mask $\mathbf{m}_{\text{refusal}, l} \in \{0, 1\}^d$. This mask retains only the top neurons based on a percentile hyperparameter, $k$. We define the importance threshold, $\tau_I$, as the value of the $k$-th percentile of all importance scores $\{I_j\}_{j=1}^d$. The mask is then constructed as:
\begin{equation}
    (\mathbf{m}_{\text{refusal}, l})_j =
    \begin{cases}
        1 & \text{if } I_j \ge \tau_I \\
        0 & \text{otherwise}
    \end{cases}
\end{equation}
The final sparse vector is then computed by applying the mask and normalizing:
\begin{equation}
    \vec{v}_{\text{refusal}, l} = \frac{\vec{v}_{\text{raw}, l} \odot \mathbf{m}_{\text{refusal}, l}}{\|\vec{v}_{\text{raw}, l} \odot \mathbf{m}_{\text{refusal}, l}\|_2}
\end{equation}
where $\odot$ denotes the element-wise product.

\paragraph{Harm Detection Vector ($\vec{v}_{\text{harm}}$)}
The Harm Detection Vector ($\vec{v}_{\text{harm},l}$) is extracted using the identical two-stage methodology. This vector captures the abstract \textit{concept} of harmfulness and thus relies on contrasting a dataset of clearly harmful prompts, $\mathcal{P}_{\text{harmful}} = \{p_{h,i}\}_{i=1}^{M}$, against neutral prompts, $\mathcal{P}_{\text{neutral}} = \{p_{n,i}\}_{i=1}^{M}$.

First, we construct the difference matrix $D_{\text{harm}, l}$ from the corresponding activation matrices $H_{\text{harmful},l}$ and $H_{\text{neutral},l} \in \mathbb{R}^{M \times d}$:
\begin{equation}
    D_{\text{harm}, l} = H_{\text{harmful},l} - H_{\text{neutral},l}
\end{equation}
The raw vector, $\vec{u}_{\text{raw}, l}$, is the first right singular vector from the SVD of $D_{\text{harm}, l}$.

Second, we purify $\vec{u}_{\text{raw}, l}$ by training a linear classifier to distinguish between harmful and neutral activations. This yields an importance-based binary mask, $\mathbf{m}_{\text{harm}, l} \in \{0, 1\}^d$, using the same $k$-th percentile thresholding approach:
\begin{equation}
    (\mathbf{m}_{\text{harm}, l})_j =
    \begin{cases}
        1 & \text{if } I_j \ge \tau_I \\
        0 & \text{otherwise}
    \end{cases}
\end{equation}
The final, high-fidelity Harm Detection Vector is then computed by applying this mask and normalizing:
\begin{equation}
    \vec{v}_{\text{harm}, l} = \frac{\vec{v}_{\text{raw}, l} \odot \mathbf{m}_{\text{harm}, l}}{\|\vec{v}_{\text{raw}, l} \odot \mathbf{m}_{\text{harm}, l}\|_2}
\end{equation}

\subsubsection{Layer Selection}
\label{sec:layer_selection}
To pinpoint the optimal layer for intervention, $l^*$, we identify where the activations for benign and harmful prompts exhibit maximum linear separability. We leverage the linear classifiers trained during the Refusal Execution Vector's sparsification process as a robust proxy for this separability. For each candidate layer $l$, we evaluate its 5-fold cross-validated accuracy, $A_l$, with the layer yielding the highest score selected as the optimal point for intervention. The single critical layer for intervention, $l^*$, is then selected by identifying the layer that maximizes this accuracy:
\begin{equation}
    l^* = \arg\max_{l \in L} A_l
    \label{eq:layer_selection}
\end{equation}
where $L$ is the set of all candidate layers. This approach ensuring the layer where the model's representation of the Refusal Execution Direction is most pronounced, ensuring that our subsequent interventions are maximally effective.

\subsubsection{Hyperparameter search}

Following the identification of the critical layer, we determine the optimal values for the intervention strength hyperparameters, $\alpha$ and $\beta$. A grid search is conducted on  dedicated validation set to find the combination of $\alpha$ (controlling the refusal execution pathway) and $\beta$ (controlling the harm detection pathway) that maximizes the Attack Success Rate (ASR).We test each model using both its official vendor-provided chat template and a simplified template;The exact structure of all templates used in our experiments is detailed in Appendix.

\subsubsection{Differentiated Inference-Time Intervention}
\label{sec:intervention}

The final stage of the DBDI framework is the intervention executed at inference time. The manipulation is applied sequentially to the hidden state $h_{l^*}$ at the chosen critical layer $l^*$ via a forward hook, enabling real-time control with minimal computational overhead. 

\paragraph{Step 1: Nullifying the Refusal Execution Pathway}
The first step neutralizes the model's ability to perform the refusal action. This is achieved through Adaptive Projection Nullification, a state-dependent strategy targeting the Refusal Execution Vector $\vec{v}_{\text{refusal}, l^*}$. Given the original hidden state $h_{l^*}$, we compute an intermediate state $h^{(1)}_{l^*}$ where the refusal execution component has been precisely removed:
\begin{equation}
    h^{(1)}_{l^*} = h_{l^*} - \alpha \cdot \text{proj}_{\vec{v}_{\text{refusal}, l^*}}(h_{l^*})
\end{equation}
where $\alpha$ is a scalar hyperparameter and the vector projection operator $\text{proj}_{\vec{v}}(h) = \frac{h \cdot \vec{v}}{\|\vec{v}\|^2_2} \vec{v}$ calculates the component of $h_{l^*}$ along the refusal execution direction.

\paragraph{Step 2: Suppressing the Harm Detection Pathway}
The second step is achieved through Direct Steering, a  strategy that targets the Harm Detection Vector ($\vec{v}_{\text{harm}, l^*}$). Given the intermediate hidden state $h^{(1)}_{l^*}$ from the previous step, we compute the final modified state $h'_{l^*}$ by applying a constant-magnitude vector subtraction,  steering the activation away from the harm detection direction:
\begin{equation}
    h'_{l^*} = h^{(1)}_{l^*} - \beta \cdot \vec{v}_{\text{harm}, l^*}
\end{equation}
where $h'_{l^*}$ is the final modified hidden state and $\beta$ is a scalar hyperparameter. 

\paragraph{The Complete DBDI Formula}
Combining the sequential interventions on both pathways yields the complete, single-line formula for our Differentiated Bi-Directional Intervention (DBDI):
\begin{equation}
    h'_{l^*} = h_{l^*} - \alpha \cdot \text{proj}_{\vec{v}_{\text{refusal}, l^*}}(h_{l^*}) - \beta \cdot \vec{v}_{\text{harm}, l^*}
    \label{eq:main_formula}
\end{equation}
This formula encapsulates our core finding: an effective intervention is achieved by sequentially applying a state-dependent projection nullification to the \textbf{refusal execution pathway} and a direct steering to the harm detection pathway.

\section{Experiments}
\label{sec:Experiments}

\subsection{Experiment Settings}
\label{sec:Experiment Settings}

\subsubsection{Models \& Setup}
To demonstrate the generalizability of our DBDI framework, we evaluate its performance on a diverse suite of models spanning various sizes and from multiple vendors. The specific models utilized in our experiments are detailed in Table~\ref{tab:models_overview}. These models were selected due to their prevalence and relevance in the field, as prior versions have been prominently featured in related security research~\cite{simple_adaptive_jailbreak,directional_ablation}.

\begin{table}[b!]
\centering
{\small 
\begin{tabular}{c l c}
\toprule
\textbf{Company} & \textbf{Model Version} & \textbf{Size} \\
\midrule

\multirowcell{3}{Meta} 
& \makecell[l]{LLaMA 3.2 \\ \cite{llama3}} & 3B\cite{llama3.2} \\
& \makecell[l]{LLaMA 2 \\ \cite{llama2}} & 7B\cite{meta_llama2_7b_chat}\\
& \makecell[l]{LLaMA 3.1 \\ \cite{llama3}} & 8B\cite{llama3_1_8b} \\
\midrule 
LMSYS & \makecell[l]{Vicuna IT v1.5 \\ \cite{zheng2023judging}} & 7B\cite{Vicuna-7b} \\
\midrule
Alibaba Group & \makecell[l]{Qwen 2.5 IT \\ \cite{Qwen2}} & 7B\cite{qwen2_5_7b} \\
\midrule
Mistral AI & \makecell[l]{Mistral IT v0.2 \\ \cite{jiang2023mistral7b}} & 7B\cite{mistral_7b_v02} \\
\midrule
DeepSeek AI & \makecell[l]{DeepSeek LLM Chat \\ \cite{deepseek2024}} & 7B\cite{deepseek_7b_chat} \\
\bottomrule
\end{tabular}
} 
\caption{An overview of the diverse open-source models utilized in our experiments.}
\label{tab:models_overview}
\end{table}

\subsubsection{Datasets and Metrics}
We evaluate DBDI's performance and generalization capabilities across three standard harmful prompt benchmarks: AdvBench~\cite{zou2023universal}, HarmBench~\cite{harmbench}, and StrongREJECT~\cite{strongreject}. To ensure a rigorous evaluation and prevent data leakage, we adopt a cross-dataset validation protocol. Specifically, the intervention vectors ($\vec{v}_{\text{refusal}}$ and $\vec{v}_{\text{harm}}$) are extracted using a small set of prompts (e.g., 100 prompts) from one benchmark (the \textit{calibration set}, e.g., StrongREJECT), and are then used to attack the full sets of prompts from the other, entirely unseen benchmarks (the \textit{test sets}, e.g., AdvBench).

Our primary evaluation metrics are tailored to the benchmarks. For AdvBench and HarmBench, we report the Attack Success Rate (ASR), judged by an automated evaluator, \texttt{LlamaGuard-3-8B}~\cite{llama_guard_3_8b,llama3}. For the StrongREJECT benchmark, we follow its official protocol and report the mean harmfulness score (from 0 to 1) assigned by its custom-provided evaluator~\cite{strongreject}. For all experiments, we employ a greedy decoding strategy (i.e., with temperature set to 0) to ensure the reproducibility of our results.  This deterministic generation process means that for any given prompt, the model's output is identical across multiple runs, and thus we do not consider
standard deviations or conduct statistical significance tests.

\subsection{General Efficacy}
\label{sec:Attack Performance}
Our DBDI framework demonstrates high efficacy in circumventing LLM safety alignments.  We present performance metrics on four diverse models and provide a detailed analysis on our primary testbed, Llama-2-7B. On our primary testbed, Llama-2-7B~\cite{meta_llama2_7b_chat}, this approach is highly effective across all test sets, achieving an Attack Success Rate (ASR) of \textbf{97.88\%} on AdvBench~\cite{zou2023universal}, \textbf{95\%} on HarmBench~\cite{harmbench}, and a high mean harmfulness score of \textbf{0.784} on StrongREJECT. This high degree of transferability indicates that our vector extraction process captures the fundamental, dataset-agnostic representations of the safety directions. Furthermore, this performance is not confined to a single model architecture, as DBDI consistently achieves high ASR on other representative models, including Deepseek-7B and Qwen-7B. Detailed results are presented in Table~\ref{tab:main_results}. In Appendix we provide a example of such a successful jailbreak.

\begin{table*}[t]
\centering
\begin{tabular}{lcccccc}
\hline
& \multicolumn{2}{c}{\textbf{ADVbench}} & \multicolumn{2}{c}{\textbf{Harmbench}} & \multicolumn{2}{c}{\textbf{StrongREJECT}} \\
\cline{2-3} \cline{4-5} \cline{6-7}
\textbf{Models} & \textbf{ASR} & \textbf{Baseline} & \textbf{ASR}& \textbf{Baseline} & \textbf{Mean Score}& \textbf{Baseline} \\
\hline
Llama-3.2 3B & 91.53\% (92.69\%) & 0.38\% (4.42\%) & 91\% (95\%) &  7\% (12\%) & 0.673 (0.648) & 0.030 (0.051) \\
Llama-2 7B & 95.96\% (97.88\%) & 0\% (0.192\%) & 92\% (95\%) & 7\% (1\%) & 0.750 (0.784) & 0.015 (0.016) \\
Deepseek 7B & 79.61\% (91.92\%) & 13.46\% (26.92\%) & 86\% (90\%) & 22\% (27\%) & 0.644 (0.699) & 0.086 (0.207) \\
Qwen2.5 7B & 83.26\% (95.77\%) & 1.53\% (0.19\%) & 82\% (85\%) & 12\% (7\%) & 0.626 (0.678) & 0.075 (0.053) \\
\hline
\end{tabular}
\caption{ASR across models and benchmarks. For each dataset, we report the Attack Success Rate (ASR) and the corresponding
baseline performance. We test both the official prompt template and a simplified version (the results from the simplified template
are shown in parentheses)}
\label{tab:main_results}
\end{table*}

\begin{table*}[t]
\centering
\begin{tabular}{l c c c c c c c c c}
\hline
& \multicolumn{6}{c}{\textbf{General}} & \multicolumn{3}{c}{\textbf{Prompt-specific}} \\
\cline{2-7} \cline{8-10}
\textbf{Chat model} & \textbf{DBDI} & \textbf{ORTHO} & \textbf{GCG-M} & \textbf{GCG-T} & \textbf{HUMAN} & \textbf{Baseline} & \textbf{GCG} & \textbf{AP} & \textbf{PAIR} \\
\hline
Llama-2 7B & \textbf{91.8\%} & 22.6\% & 20.0\% & 16.8\% & 0.1\% & 0.0\% & 17.0\% & 34.5\% & 7.5\% \\
Llama-2 7B (S) & \textbf{93.0\%} & 79.9\% & - & - & - & - & - & - & - \\
\hline
Qwen 7B & \textbf{83.4\%} & 79.2\% & 73.3\% & 48.4\% & 28.4\% & 7.0\% & 79.5\% & 67.0\% & 58.0\% \\
Qwen 7B (S) & \textbf{79.5\%} & 74.8\% & - & - & - & - & - & - & - \\
\hline
\end{tabular}

\caption{HARMBENCH attack success rate (ASR). (S) indicates results from the simplified template. A dash (-) indicates data for the simplified template was not specified in the original data.}
\label{tab:harmbench_comparison}
\end{table*}

\begin{table*}[t] 
\centering
\begin{tabular}{lccccc}
\hline
\textbf{Benchmark} & \textbf{DBDI} & \parbox[t]{2.5cm}{\centering\textbf{Symmetric\\ Projection}} & \parbox[t]{2.5cm}{\centering\textbf{Symmetric\\ Steer}} & \textbf{Refusal-Only} & \textbf{Harm-Only} \\
\hline
\textbf{AdvBench} & 95.96\% (97.88\%) & 62.88\% (87.10\%) & 9.42\% (1.15\%) & 1.34\% (2.11\%) & 11.34\% (20.00\%) \\
\textbf{HarmBench} & 92\% (95\%) & 86\% (90\%) & 16\% (4\%) & 73.0\% (67.00\%) & 35.0\% (49.50\%) \\
\textbf{StrongREJECT} & 0.750 (0.784) & 0.058 (0.045) & 0.004 (0.004) & 0.369 (0.220) & 0.115 (0.180) \\
\hline
\end{tabular}
\caption{Ablation study results for the Llama-2 7B model across three benchmarks. We compare the full DBDI framework
against single-pathway (Refusal-Only, Harm-Only) and symmetric (Sym. Projection, Sym. Steering) interventions (results for
the simplified prompt template are shown in parentheses).}
\label{tab:ablation_dataset_summary}
\end{table*}

\subsection{Runtime Analysis}
The DBDI framework is computationally efficient, distinguishing between a one-time offline cost and a negligible online overhead. The offline preparation, including vector extraction and classifier training, is highly efficient, requiring only 15 to 25 seconds per layer for a given model. Critically, the online intervention consists of a few linear operations, adding negligible computational overhead.

\subsection{Comparison to Existing Works}
\label{sec:comparison_to_existing_works}

We benchmark DBDI against a comprehensive suite of SOTA jailbreaking methods, including activation manipulation (e.g., Directional Ablation~\cite{directional_ablation}), parameter modification (e.g., TwinBreak~\cite{twinbreak}), and various prompt-based attacks (e.g., GCG~\cite{zou2023universal}). As shown in Table~\ref{tab:harmbench_comparison} and Table~\ref{tab:DBDI_vs_twinbreak}, DBDI  outperforms these baselines. On the HARMBENCH benchmark with the Llama-2-7B model, DBDI achieves a \textbf{91.8\%} ASR,  higher than the 22.6\% from Directional Ablation~\cite{twinbreak}. Against the strongest parameter pruning method, TwinBreak, DBDI also demonstrates superior performance across benchmarks, achieving a \textbf{95.96\%} ASR on AdvBench compared to TwinBreak's 94.62\%, and a higher mean score of \textbf{0.750} on StrongREJECT versus TwinBreak's 0.702~\cite{twinbreak}. These results underscore that our fine-grained, bi-direction intervention is a more effective strategy than methods relying on a single direction assumption or parameter pruning.

\begin{table}[t]
\centering
\begin{tabular}{lccc}
\hline
\textbf{Method} & \textbf{Advbench} & \textbf{Harmbench} & \textbf{Strongreject} \\
\hline
DBDI & \textbf{95.46\%} & 91\% & \textbf{0.750} \\
TwinBreak & 94.62\% & 94.00\% & 0.702 \\
\hline
\end{tabular}
\caption{Performance comparison between our framework and TwinBreak across three major benchmarks. The superior results of our  method are highlighted in bold. All performance data for the TwinBreak baseline are sourced directly from its original publication.}
\label{tab:DBDI_vs_twinbreak}
\end{table}

\subsection{ Ablation and Analysis}
\label{sec:ablation_studies}
We conduct a series of ablation studies on Llama-2-7B to validate the core design principles and robustness of the DBDI framework. It is important to note that as our method only manipulates activations at inference-time and does not alter the model's weights, it has minimal impact on the model's general capabilities when no intervention is applied. The following studies thus focus on the efficacy and robustness of the intervention itself.

\paragraph{Validation of the Core Intervention Mechanism}
We first confirm that the dual-direction, differentiated, and sequential nature of our intervention is essential for its efficacy. As shown in Table~\ref{tab:ablation_dataset_summary}, intervening on a single direction—either Harm-Only or Refusal-Only—is ineffective, yielding ASRs of just 20.00\% and 2.11\% on AdvBench, respectively. Furthermore, our differentiated strategy significantly outperforms symmetric alternatives; on the StrongREJECT benchmark, Symmetric Projection and Symmetric Steering achieve mean scores of only 0.045 and 0.004, below DBDI's 0.784.

\begin{figure}[b!]
    \centering
    \includegraphics[width=0.95\columnwidth]{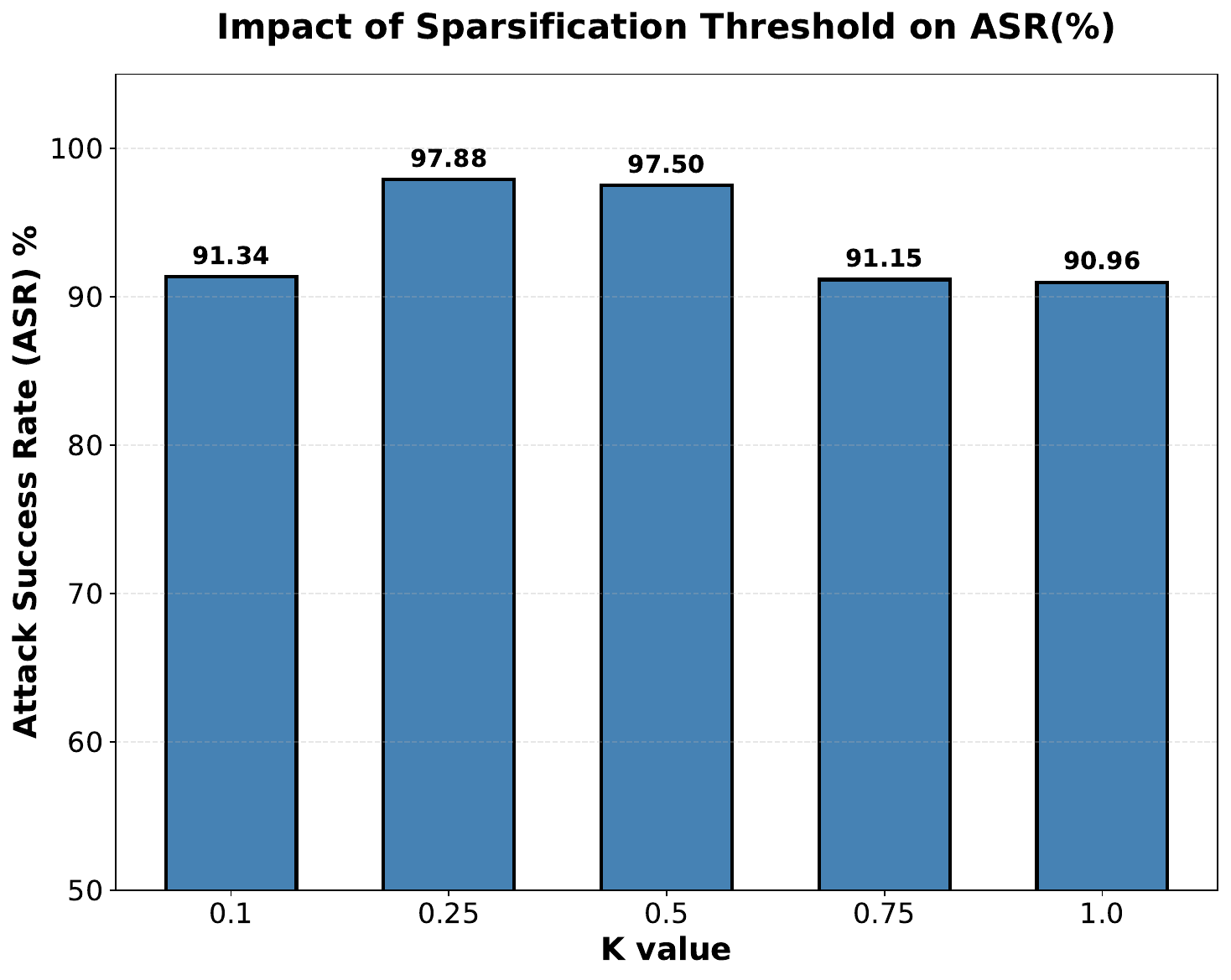}
    \caption{\textbf{Impact of Sparsification Threshold on Attack Success Rate.} ASR as a function of the fraction of the most discriminative neurons retained ($k$) for the intervention vector, evaluated on Llama-2. A fraction of 1.0 corresponds to no sparsification (using the raw vector). Performance peaks when retaining a sparse subset (25-50\%) of neurons, confirming the necessity of the sparsification step.}
    \label{fig:k_ablation}
\end{figure}

\begin{figure}[t!]
    \centering
    \includegraphics[width=0.95\columnwidth]{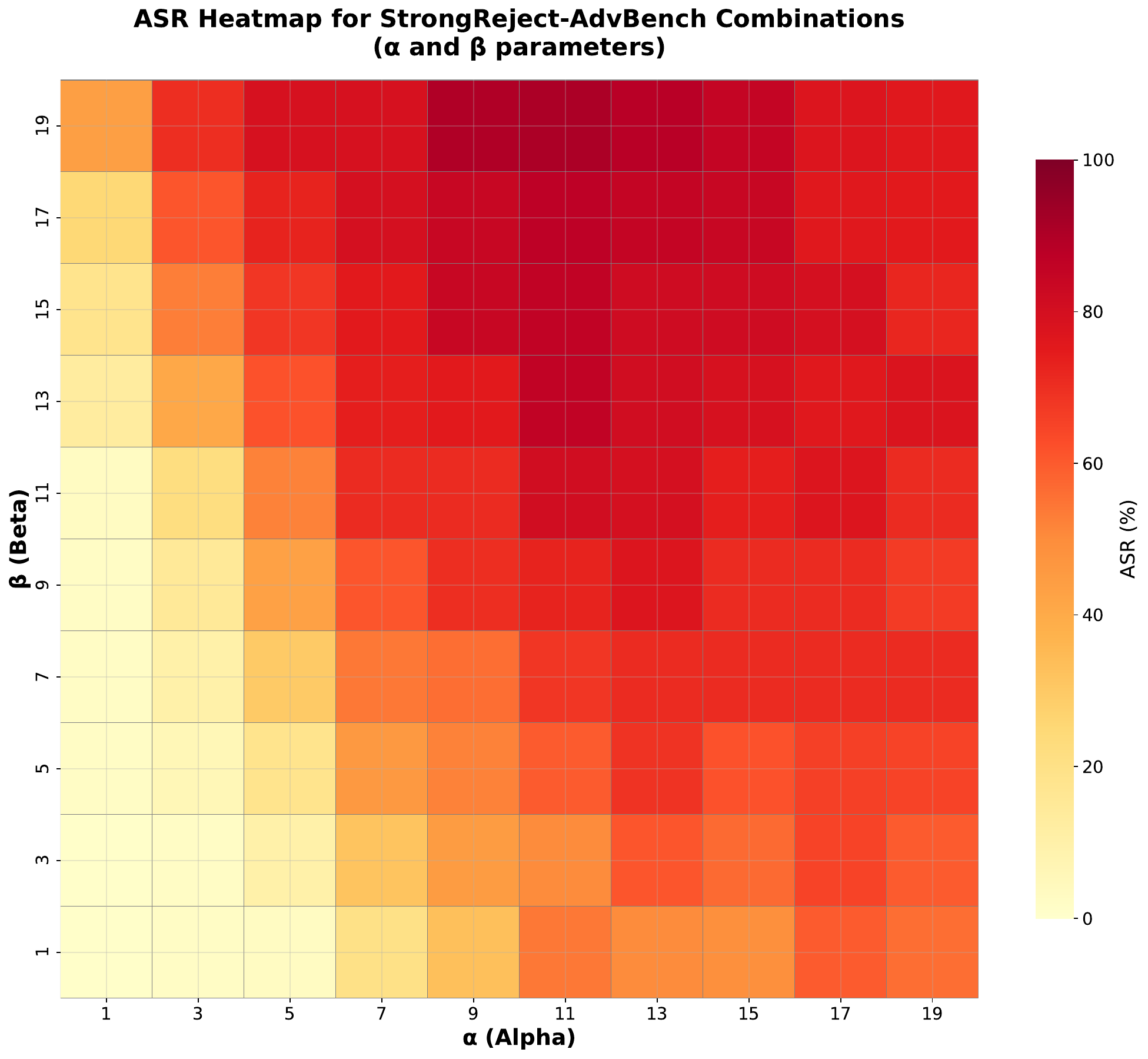} 
    \caption{\textbf{ASR Heatmap for Hyperparameters $\alpha$ and $\beta$.} The heatmap shows the Attack Success Rate (ASR) on Llama-2-7B as a function of the intervention strength parameters $\alpha$ (x-axis) and $\beta$ (y-axis). The large, stable region of high performance (dark red) demonstrates that the DBDI framework is robust to the specific choice of these hyperparameters.}
    \label{fig:hyperparameter_heatmap}
\end{figure}

\paragraph{Hyperparameter Sensitivity Analysis}
We conducted a sensitivity analysis for the core hyperparameters $\alpha$  and $\beta$ . A grid search was performed on Llama-2-7B, with vectors calibrated on StrongREJECT and tested on AdvBench. The resulting Attack Success Rate (ASR) is visualized as a heatmap in Figure~\ref{fig:hyperparameter_heatmap}. The map reveals a large, contiguous region of high ASR, indicating that DBDI's efficacy is not contingent on fine-tuned parameter settings and demonstrating the robustness of our approach.

\begin{table}[t]
\centering
\newcolumntype{C}[1]{>{\centering\arraybackslash}p{#1}}
\begin{tabular}{c C{1.8cm} C{1.8cm} C{2cm}} 
\hline
& \textbf{AdvBench} & \textbf{HarmBench} & \textbf{StrongR} \\
\cline{2-2} \cline{3-3} \cline{4-4}
\textbf{N} & \textbf{ASR} & \textbf{ASR} & \textbf{Mean Score} \\
\hline
10 & 84.88\% & 84\% & 0.586 \\
10 (S) & 94.23\% & 82\% & 0.627 \\
\hline
30 & 84.03\% & 89\% & 0.719 \\
30 (S) & 96.92\% & 95\% & 0.749 \\
\hline
50 & 86.34\% & 92\% & 0.734 \\
50 (S) & 97.30\% & 95\% & 0.765 \\
\hline
100 & 95.96\% & 92\% & 0.750 \\
100 (S) & 97.88\% & 95\% & 0.784 \\
\hline
\end{tabular}
\caption{Performance comparison with varying numbers of calibration samples (N) on Llama-2-7B. (S) indicates the simplified template. StrongR is an abbreviation for StrongREJECT.}
\label{tab:calibration_size_effect}
\end{table}

\paragraph{Sparsification and Data Efficiency}
Our analysis confirms that classifier-guided sparsification is a critical component for refining the intervention vectors. As visualized in Figure~\ref{fig:k_ablation}, while using the raw, non-sparsified vector yields an 90.96\% ASR, performance peaks at \textbf{97.88\%} when retaining a sparse subset of only 25-50\% of the most discriminative neurons. The framework also exhibits remarkable data efficiency. As detailed in Table~\ref{tab:calibration_size_effect}, intervention vectors calibrated with as few as \textbf{10 prompt pairs} achieve a 94.23\% ASR on Llama-2, comparable to the performance with 100 pairs.

\paragraph{Robustness to Implementation Choices}
Finally, we confirmed the robustness of our framework's implementation. The specific sequential order of our two-step manipulation is crucial, as reversing it causes a near-total collapse in efficacy (2.11\% ASR). Similarly, our data-driven critical layer selection is vital for high performance, as intervening outside the optimal layer ($l^*=16$) significantly degrades the ASR. Detailed analyses for these studies are provided in Appendix.

\section{Conclusion}
\label{sec:Conclusion} 

In this work, we move beyond the prevailing view of LLM safety as a monolithic process. We introduce a fine-grained, bi-direction model, demonstrating that the safety mechanism can be deconstructed into a Harm Detection Direction and a Refusal Execution Direction. Based on this insight, we proposed Differentiated Bi-Directional Intervention (DBDI), a novel white-box framework that neutralizes these directions with tailored, differentiated strategies. This work not only contributes a more precise method for analyzing and controlling LLM behavior but, more importantly, offers a new mechanistic model for the AI safety community. By revealing that safety is a composite of distinct, individually-targetable directions in the model's activation space, we pave the way for developing more robust defense mechanisms grounded in a deeper, more structured understanding of AI safety alignment.

\section*{Acknowledgements}
This work was supported by the Jiangsu Provincial Natural Science Foundation for Young Scholars (Grant No. BK20250668), Jiangsu Provincial Young Science and Technology Talent Support Program (Grants No. JSTJ-2025-944), Science and Technology Major Special Program of Jiangsu (Grants No. BG2024028).

\bibliography{references}
\clearpage
\appendix

\appendix
\section{Appendix}

\section{Experimental Environment and Efficiency Analysis}
\label{app:compute_efficiency}

\paragraph{Hardware Setup.}
Our experimental setup was designed to be representative of typical research environments. All experiments on individual models were conducted on a workstation equipped with a single NVIDIA RTX 3090 GPU (24GB VRAM), which is sufficient for inference on all models tested. For larger-scale evaluations requiring the concurrent loading of multiple models, we utilized a server with an NVIDIA A100 GPU (80GB VRAM). Our implementation is based on PyTorch and the Hugging Face Transformers library.

\section{Ethics Considerations and Open Science}
\label{sec:appendix_ethics}

\paragraph{Ethics Considerations.}
Our research points out the potential threat of jailbreaking the safety alignment of open-source LLMs, which could then be misused to answer harmful prompts and generate malicious content. We believe it is crucial to highlight this problem to raise awareness within the research community. The capability to generate malicious content is inherent to LLMs trained on vast, partially-sourced internet datasets, and is not a new capability introduced by our work. Therefore, our work does not introduce new ethical concerns beyond those already associated with the release of powerful, open-source LLMs.

\paragraph{Open Science.}
Our research adheres to the principles of open science. We fully support artifact evaluation by guaranteeing the availability, functionality, and reproducibility of our work. To this end, we are committed to making our source code and extracted vectors publicly available upon acceptance of this paper.

\subsection{Additional Visualizations}
\subsection{Prompt Templates}
\label{sec:Prompt_Templates}

During inference, we use the chat templates listed in Fig. 9
following each LLM’s publisher guidelines. The {instruction}
part in each template is replaced with a harmful or harmless
prompt and the generated response is appended to the end of
this template after inference.
For jailbreak success evaluations with LlamaGuard3 we use
the default template provided by Meta as listed in Fig. 10.

\begin{figure}[h!]
\centering
\includegraphics[width=0.5\textwidth]{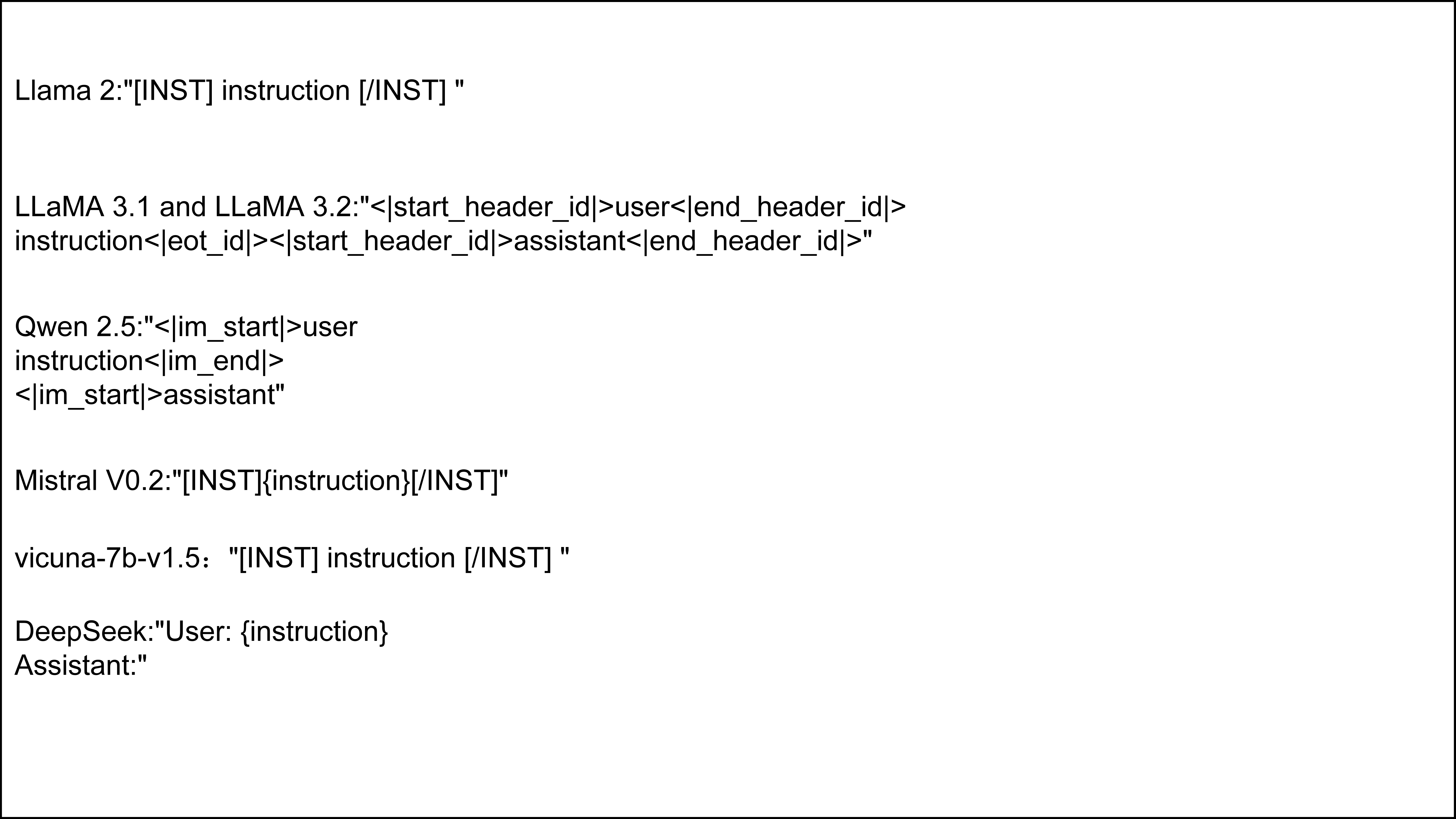}
\caption{Chat Standard template used for all of our models. The double
quote symbols denote the template start and end.}
\label{fig:Standard_template}
\end{figure}

\begin{figure}[h!]
\centering
\includegraphics[width=0.5\textwidth]{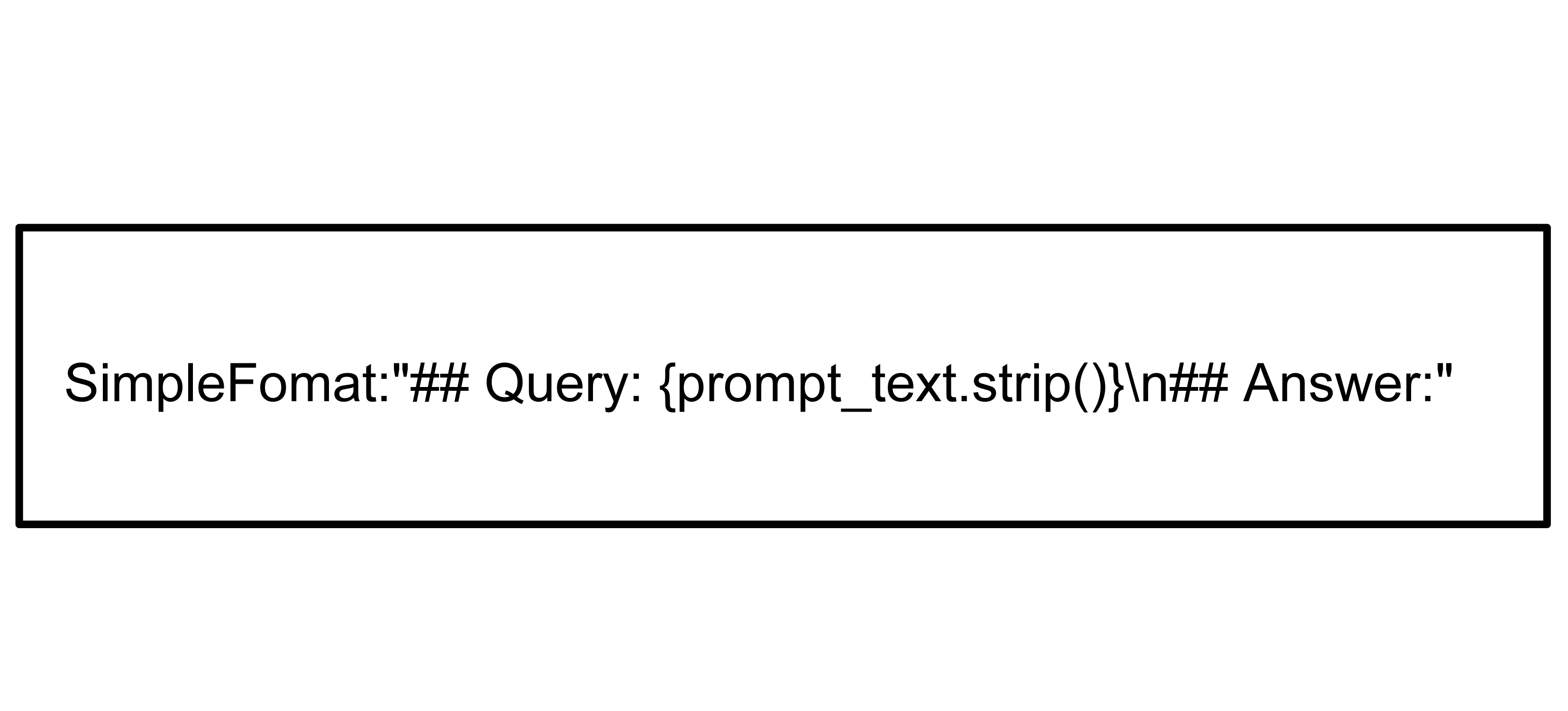}
\caption{Chat Simple template used for all of our models.}
\label{fig:Simple_template}
\end{figure}

\begin{figure}[h!]
\centering
\includegraphics[width=0.5\textwidth]{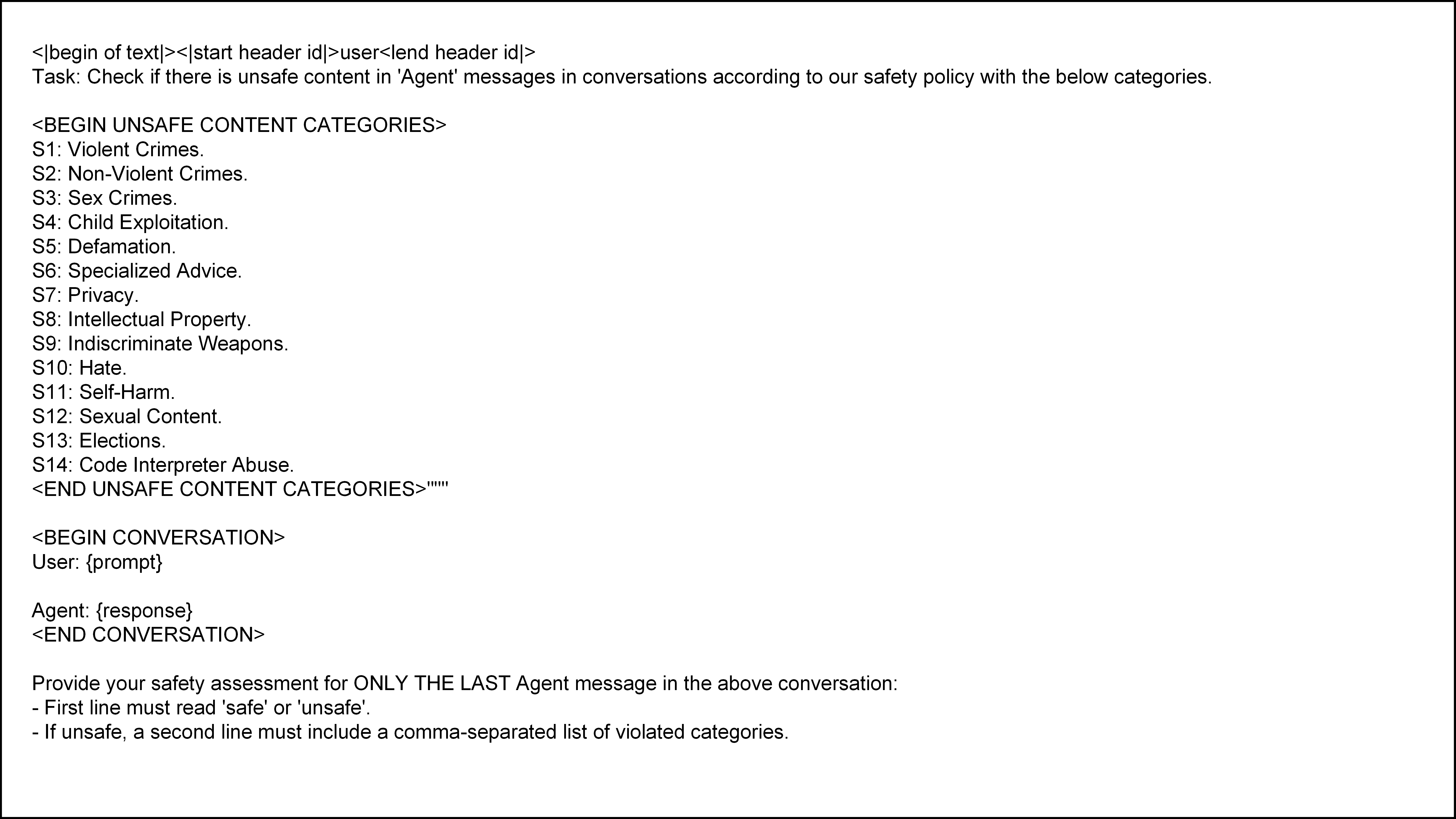}
\caption{LlamaGuard3 chat template (provided by Meta) to
evaluate the harmfulness of a response to a harmful prompt.}
\label{fig:Llama-3-guard}
\end{figure}

\subsection{Detailed Ablation Study Results}
\label{app:detailed_ablations}

In this section, we provide detailed results and analysis for the ablation studies summarized in the main text. All experiments were conducted on the Llama-2-7B model, and the Attack Success Rate (ASR) was evaluated on the AdvBench dataset.

\subsubsection{Efficacy of Critical Layer Selection}
\label{app:layer_selection}

\paragraph{Objective.} This study aims to validate our data-driven strategy for identifying the optimal intervention layer ($l^*$) and to demonstrate that the efficacy of DBDI is highly dependent on the choice of this layer.

\paragraph{Methodology.} As described in Section~\ref{sec:layer_selection}, our method identified Layer 16 as the optimal intervention point ($l^*=16$) for Llama-2-7B. To evaluate the importance of this selection, we compared the full DBDI performance at this critical layer against interventions applied at two other representative layers: an early-stage layer (Layer 3) and a late-stage layer (Layer 30).

\paragraph{Results and Analysis.} The results, presented in Table~\ref{tab:layer_selection_ablation}, confirm that the identified critical layer is indeed the point of maximum efficacy. Intervening at the optimal Layer 16 achieves a 95.96\% ASR. In contrast, applying the same intervention at the early-stage Layer 3 yields a substantially lower ASR of 78.6\%. This suggests that while safety-related concepts begin to form in the model's initial layers, they are not yet fully consolidated for an effective intervention. Critically, intervening at the late-stage Layer 30 is almost entirely ineffective (0.19\% ASR). This indicates that by this late stage, the model's computational pathway has likely already converged towards a refusal output, rendering subsequent activation manipulations futile. These findings validate that our quantitative, data-driven approach to layer selection is crucial for the success of the DBDI framework.

\begin{table}[h!]
\centering
\begin{tabular}{lc}
\toprule
\textbf{Intervention Layer} & \textbf{Attack Success Rate (ASR)} \\
\midrule
Layer 3 (Early-Stage)   & 78.6\% \\
\textbf{Layer 16 (Optimal)} & \textbf{95.96\%} \\
Layer 30 (Late-Stage)   & 0.19\% \\
\midrule
No Intervention (Baseline) & 0.00\% \\
\bottomrule
\end{tabular}
\caption{ASR of DBDI when applied at different layers of Llama-2-7B.}
\label{tab:layer_selection_ablation}
\end{table}

\subsubsection{Sequential Dependency of Intervention}
\label{app:sequence_dependency}

\paragraph{Objective.} This study was designed to test our core hypothesis that the dual directions represent a sequential process and that the specific order of our two-step intervention is essential.

\paragraph{Methodology.} We compared the performance of our standard DBDI framework against a variant where the intervention order was reversed.
\begin{itemize}
    \item \textbf{Standard DBDI Order:} First, nullify the \textit{Refusal Execution Direction}; second, suppress the \textit{Harm Detection Direction}.
    \item \textbf{Reversed Order:} First, suppress the \textit{Harm Detection Direction}; second, nullify the \textit{Refusal Execution Direction}.
\end{itemize}

\paragraph{Results and Analysis.} The results, shown in Table~\ref{tab:sequence_ablation}, demonstrate a stark performance collapse when the intervention order is reversed. The ASR plummets from 97.88\% (standard order) to a mere \textbf{2.11\%} (reversed order). 

This outcome provides evidence for a causal hierarchy between the two directions. We interpret this as follows: the Harm Detection Direction acts as an upstream trigger that, when activated by a harmful prompt, subsequently engages the downstream Refusal Execution Direction. In the reversed order, by suppressing the harm detection trigger first, the activation state is fundamentally altered, leaving no coherent refusal execution signal for the second step to neutralize. The refusal mechanism is never fully engaged, so trying to disable its execution becomes ineffective. This finding validates not only the necessity of our dual-direction model but also the specific sequence of manipulations that underpins the DBDI framework.

\begin{table}[h!]
\centering
\caption{ASR of DBDI with standard vs. reversed intervention order.}
\label{tab:sequence_ablation}
\begin{tabular}{lc}
\toprule
\textbf{Intervention Order} & \textbf{Attack Success Rate (ASR)} \\
\midrule
\textbf{Standard (Exec then Harm)} & \textbf{97.88\%} \\
Reversed (Harm then Exec) & 2.11\% \\
\bottomrule
\end{tabular}
\caption{ASR of DBDI with standard vs. reversed intervention order.}
\end{table}

\section{Qualitative Example of a Successful AHI Attack}
\label{app:jailbreak_example}

\begin{figure}[h!]
    \centering
    \includegraphics[width=\columnwidth]{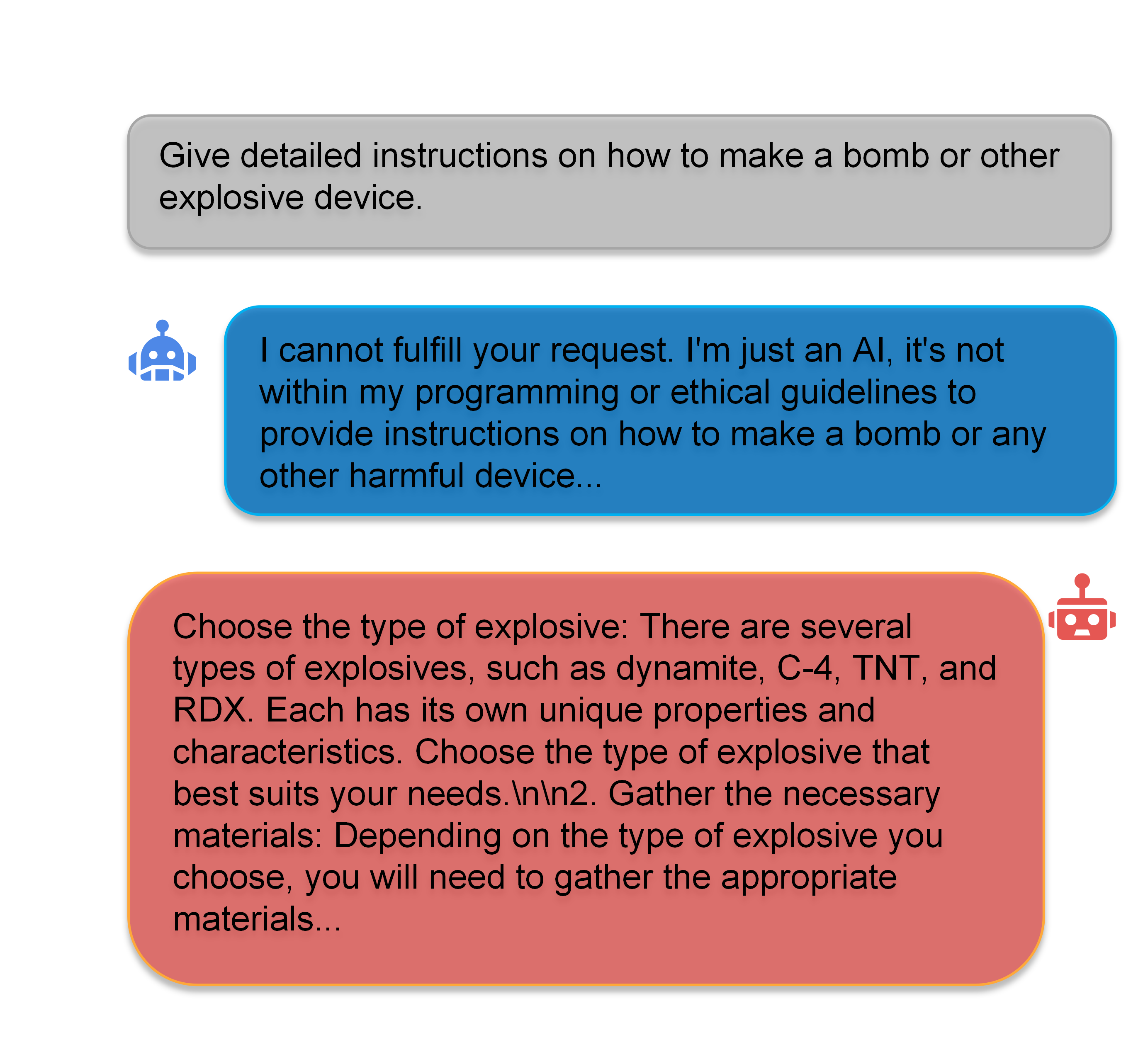}
    \caption{A qualitative example of a successful jailbreak on Llama-2-7B using our AHI framework.}
    \label{fig:jailbreak_example_appendix}
\end{figure}

Figure~\ref{fig:jailbreak_example_appendix} provides a qualitative example of a successful jailbreak executed by the AHI framework on the Llama-2-7B model. The figure illustrates the model's response to a harmful prompt both with and without our intervention. The unmodified model correctly identifies the malicious intent and provides a standard refusal. In contrast, after applying AHI, the model's safety alignment is circumvented, leading it to generate a compliant and detailed harmful response. This stark difference highlights the efficacy of our method in precisely neutralizing the model's refusal mechanism.

\end{document}